\documentclass[twocolumn,showpacs,aps,prl,superscriptaddress]{revtex4}

\def\Dtwostm  {\ensuremath{D_2^\ast(2460)^-}\xspace}
\def\Kstz    {\ensuremath{K^\ast(892)^0}\xspace}
\newcommand{\DE}{\ensuremath{\Delta E}\xspace}
\newcommand{\mES}{\ensuremath{m_\mathrm{ES}}\xspace}
\newcommand{\thetaC}{\ensuremath{\theta_C}\xspace}

\usepackage{graphicx}
\usepackage{dcolumn}
\usepackage{amsmath}
\usepackage{epsfig}
\usepackage{bigstrut}

\input babarsym.tex

\newcommand{\BABARPubYear}    {05}
\newcommand{\BABARPubNumber}  {43}
\newcommand{\SLACPubNumber} {11474}
\newcommand{\LANLNumber} {0509036}

\def\figurebox#1#2#3{%
    \def\arg{#3}%
    \ifx\arg\empty
    {\hfill\vbox{\hsize#2\hrule\hbox to #2{\vrule\hfill\vbox to #1{\hsize#2\vfill}\vrule}\hrule}\hfill}%
    \else
    {\hfill\epsfbox{#3}\hfill}%
    \fi}

\begin{document}

\preprint{\babar-PUB-\BABARPubYear/\BABARPubNumber} 
\preprint{SLAC-PUB-\SLACPubNumber} 

\begin{flushleft}
\babar-PUB-\BABARPubYear/\BABARPubNumber\\
SLAC-PUB-\SLACPubNumber\\
hep-ex/\LANLNumber\\[5mm]
\end{flushleft}

\title{
{\large \bf
Measurement of Branching Fractions and Resonance Contributions \\ for
$\Bz\ra\Dzb\Kp\pim$ and Search for $\Bz\ra\Dz\Kp\pim$ Decays} 
}

%
\author{B.~Aubert}
\author{R.~Barate}
\author{D.~Boutigny}
\author{F.~Couderc}
\author{Y.~Karyotakis}
\author{J.~P.~Lees}
\author{V.~Poireau}
\author{V.~Tisserand}
\author{A.~Zghiche}
\affiliation{Laboratoire de Physique des Particules, F-74941 Annecy-le-Vieux, France }
\author{E.~Grauges}
\affiliation{IFAE, Universitat Autonoma de Barcelona, E-08193 Bellaterra, Barcelona, Spain }
% ...
\collaboration{The \babar\ Collaboration}
\noaffiliation

\date{\today}

\begin{abstract}
Using 226 million $\Upsilon(4S)\ra\BB$ events collected with the \babar\
detector at the PEP-II \epem storage ring
at the Stanford Linear Accelerator Center,
we measure the branching fraction
for $\Bz\ra\Dzb\Kp\pim$, excluding $\Bz\ra \Dstarm\Kp$, to be
$
{\cal B}(\Bz\ra\Dzb\Kp\pim) = (88 \pm 15 \pm 9 )\times 10^{-6}\;.
$
We observe $\Bz\ra\Dzb\Kstz$ and $\Bz\ra\Dtwostm\Kp$
contributions. 
The ratio of branching fractions ${\cal B}(\Bz\ra\Dstarm\Kp)/{\cal
  B}(\Bz\ra\Dstarm\pip)= (7.76\pm 0.34 \pm 0.29)\%$ is measured
separately.
The branching fraction for the suppressed mode $\Bz\ra\Dz\Kp\pim$ is 
$
{\cal B}(\Bz\ra\Dz\Kp\pim) < 19 \times 10^{-6}\,
$
at the 90\% confidence level.

\end{abstract}


\maketitle


A theoretically clean method for measuring the angle $\gamma =
\mathrm{arg}(-V_{ud}V^*_{ub} / V_{cd}V^*_{cb})$ in the
unitarity triangle of the
Cabibbo-Kobayashi-Maskawa (CKM) quark-mixing
matrix~\cite{CKM} in the Standard Model of particle physics
utilizes decay modes of the type $\B\ra D K$. 
Several methods have been proposed~\cite{GW,ADS,GGS} to extract
$\gamma$ from these decays using interference effects
between $\b\ra\u \cbar\s$ and $\b\ra {\c} {\ubar} \s$ processes. However,
the $\b\ra\u \cbar\s$ amplitude is suppressed by a color factor in
addition to the CKM factor $|V_{ub}V^*_{cs}/V_{cb}V^*_{us}| \simeq 0.4$,
and the extraction of $\gamma$ with methods in Ref.~\cite{GW,ADS}
 is subject to an
eight-fold ambiguity due to unknown strong phases.

Three-body $B\ra D K \pi$ decays have been
proposed~\cite{APS,Gronau:2002mu} as an 
alternative method for measuring $\gamma$.
In these modes, the CKM-suppressed $\b\ra\u \cbar\s$ processes include
color-allowed diagrams; thus
larger decay rates and more significant \CP violation effects are
possible. 
In addition, a $D K \pi$ Dalitz plot analysis can resolve the strong phase and
reduce the ambiguity to two-fold, similar to Ref.~\cite{GGS}.
The sensitivity to $\gamma$ in these decays is determined by the size
of the overlapping $\b\ra {\c} {\ubar} \s$ and $\b\ra\u \cbar\s$ amplitudes
in the Dalitz plot. 

In this Letter, we report the measurements of the branching fraction
for the CKM-favored $\Bz\ra\Dzb\Kp\pim$~\cite{footnote1}
decay and dominant resonance contributions,
and the search for the CKM-suppressed $\Bz\ra\Dz\Kp\pim$ decays.
The flavor of the \B meson is tagged by the charge of
the prompt kaon. 
The favored mode has been previously observed through its dominant
resonances $\Dstarm\Kp$~\cite{Abe:2001wa} 
and
$\Dzb\Kstz$~\cite{Krokovny:2002ua}.
Since $\Dstarm\Kp$
occupies only a very small region of the allowed phase space, we treat
it separately 
and measure the ratio 
$r= {\cal B}(\Bz\ra\Dstarm\Kp)/{\cal B}(\Bz\ra\Dstarm\pip)$,
which can be used to test factorization and flavor-SU(3) symmetry.

Signal events are selected from 226 million \BB pairs collected with
 the \babar\
detector~\cite{Aubert:2001tu} at the PEP-II asymmetric-energy storage ring.
Charged tracks are detected by a five-layer silicon
vertex tracker and a 40-layer drift chamber. Hadrons are identified
based on the ionization energy loss in the tracking system and the
opening angle of the Cherenkov radiation in a ring-image detector~\cite{DIRC}.
Photons are measured by an electromagnetic calorimeter. These
systems are mounted inside a 1.5-T solenoidal super-conducting magnet.

The \Dz candidate is reconstructed through $\Km\pip$, $\Km\pip\piz$,
and $\Km\pip\pim\pip$ channels, where the measured invariant mass is
required to be within 20, 35, and 20~\mevcc, respectively, of the
nominal \Dz mass~\cite{PDG}, corresponding to 3.0, 2.5 and
3.0~$\sigma$. A vertex fit is performed with the mass 
constrained to the nominal value. The \piz candidate is formed from
two photon candidates with invariant mass between $115$ and
$150~\mevcc$. 

For the measurement of the ratio $r$, the
\Dz is combined with a low momentum $\pi$ to
form a \Dstar candidate, with its vertex constrained to the interaction
point (beam spot). Candidates with mass difference
$m_{\Dz\pi}-m_{\Dz}$ between 144 and 147~\mevcc are retained.
A charged track, assumed to have the pion mass, is combined
with the \Dstar to form a \Bz candidate.
The $\chi^2$ probabilities for both the \Dstar and \Bz vertex fits are required
to be greater than 0.1\%.
To reject jet-like continuum background, the normalized Fox-Wolfram second
moment $R_2$~\cite{FW}, computed with charged tracks and neutral clusters,
is required to be less than 0.5, and $|\cos\theta_T|$ less than $0.85$
where $\theta_T$ is the thrust angle between the \Bz candidate and the
rest of the event in the \epem center-of-mass (CM) frame. 

For $\Bz\ra\Dzb\Kp\pim$ and $\Dz\Kp\pim$
measurements, the \Bz candidate is formed by combining a \Dz
candidate with oppositely charged pion and kaon candidates.
We select candidates outside the $\Dstarm\Kp$ region
($142.5 < m_{\Dz\pi}-m_{\Dz} < 148.5\mevcc $, a $6\sigma$ window).
The measured \Dz invariant mass
must be within $12$, $28$, and $8.5~\mevcc$ of the nominal
\Dz mass for $K\pi$, $K\pi\piz$, and $K\pi\pi\pi$ modes, respectively.
Candidates are rejected if the $\Dz\ra K\pi\piz$ decay probability,
computed with the Dalitz parameters measured in Ref.~\cite{kpipi0Dalitz}, is
less than 6\% of the maximum value.
The $\chi^2$ probability of the \Dz (\Bz) vertex fit is required
to be greater than 0.5\% (2\%).
All charged tracks are required to have at least 12
hits in the drift chamber and transverse momentum greater than $100~\mevc$. 
Both kaon candidates are required to be consistent with the kaon
hypothesis.
Prompt pion candidates consistent with the kaon hypothesis are rejected.

To further reduce  the continuum background, 
$|\cos\theta_B^*|$ must be less than 0.9, where $\theta_B^*$ is the
polar angle of the \Bz candidate in the CM frame.
A Fisher discriminant $\cal F$ is
formed based on $R_2$, $\cos\theta_T$, $\theta_B^*$, and
two moments $L_{0}$ and $L_{2}$, where $L_{i}= \sum_j p^*_j
|\cos\theta^*_j|^i$, summed over the remaining particles $j$ in the
event, where $\theta^*_j$ and $p^*_j$  are the 
angle with respect to the \Bz thrust and the momentum in the CM frame.
Different cuts on $\cal F$ are applied for each mode to optimize the
signal significance based on simulated event samples.
Candidates used in the subsequent fits have beam-energy
substituted mass $\mES= \sqrt{(\sqrt{s}/2)^2 - (p^*)^2} >
5.2~\gevcc$ and energy difference $|\DE|= |E^*- \sqrt{s}/2| < 150~\mev$,
where $E^*$ and $p^*$ are the energy and momentum of the
\Bz candidate and $\sqrt{s}$ is the total energy in the CM frame.

We study five samples separately: 
(a) $\Bz\ra\Dzb\Kp\pim$ excluding the
$\Dstarm\Kp$ contribution,
(b) $\Bz\ra\Dz\Kp\pim$,
(c) $\Bz\ra\Dzb\Kstz$,
(d) $\Bz\ra\Dtwostm\Kp$, and
(e) $\Bz\ra\Dstarm h^+$ where $h^+$ is a pion or kaon.
Samples (c) and (d) are subsets of (a), where the resonances are selected
within 1.5 times their full widths~\cite{PDG}.

\begin{figure}[tb]
\begin{center}
  \includegraphics[width=0.48\textwidth]{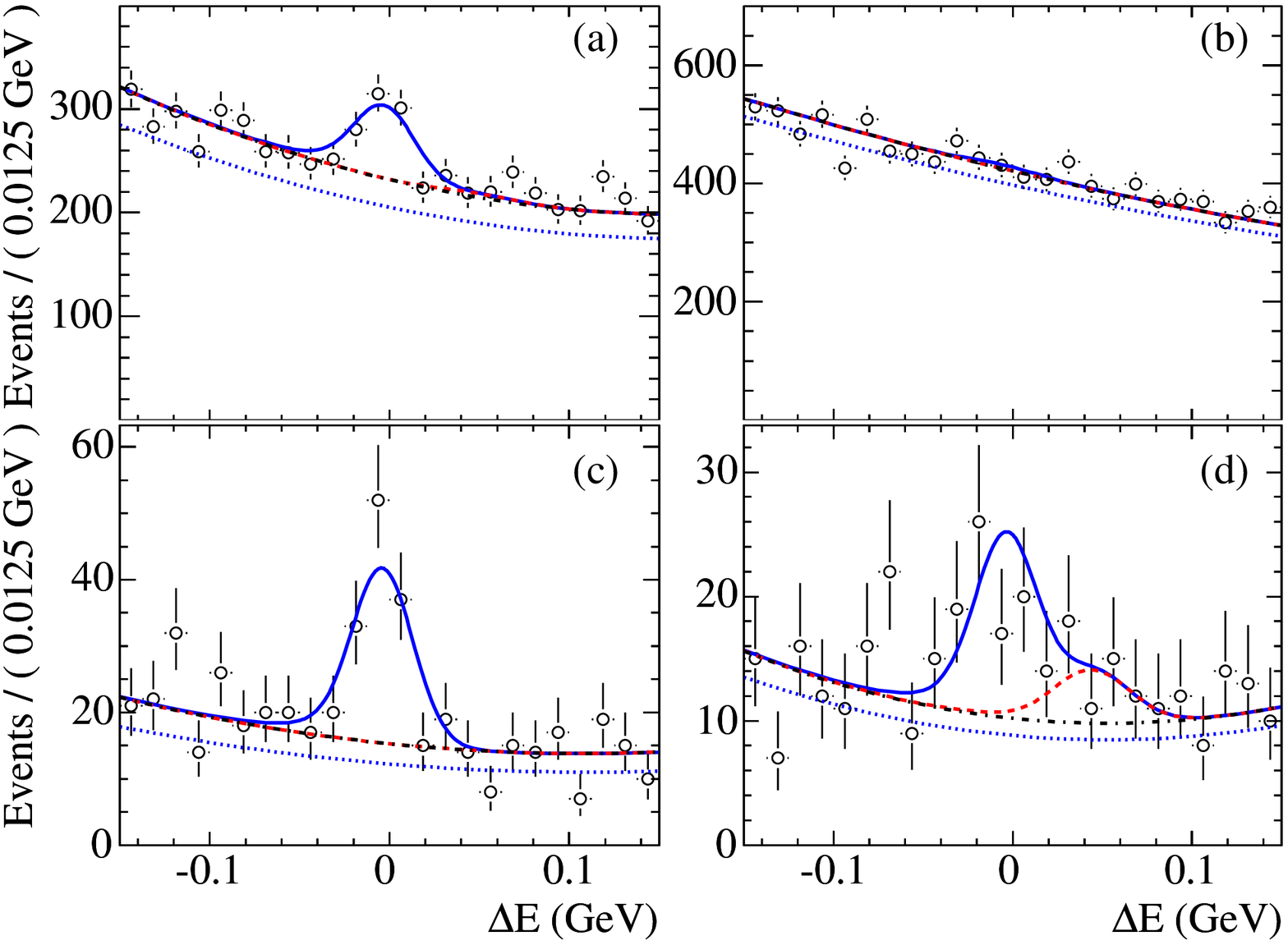}
\end{center}
\caption{\DE distributions 
  and  PDF projections with 
  $\mES>5.27~\gevcc$ for (a) $\Bz\ra\Dzb\Kp\pim$ excluding
  $\Dstarm\Kp$ candidates, (b) 
  $\Bz\ra\Dz\Kp\pim$, (c) $\Bz\ra\Dzb\Kstz$ and (d)
  $\Bz\ra\Dtwostm\Kp$, for the
  three \Dz modes combined. 
  Circles with error bars are data points.
  Four curves from top to bottom represent: 
  the total PDF (solid), total background (dashed), combinatorial
  background plus peaking background B described in the text
  (dot-dashed) and combinatorial background only (dotted). In
  (a)--(c), the middle two curves overlap 
  because the peaking background A is negligible.}
\label{fig:de}
\end{figure}

For samples (a)--(d), 
a two-dimensional (\mES, \DE) unbinned-maximum-likelihood fit is used
to determine the signal yields.
The signal component is the product of a
Gaussian in \mES centered at the \Bz mass and a Crystal Ball
lineshape~\cite{crball} in \DE centered near zero. The combinatorial
background component is modeled with an Argus threshold
function~\cite{argus} in \mES and a second-order polynomial in \DE.
Two background components peak in \mES: peaking background A
describes the $\Bz\ra D^{**-}\pip$ contribution, which also peaks in
\DE but the peak is shifted by about $+50~\mev$ because the pion is
misidentified as a kaon; peaking background B uses a second-order
polynomial in \DE to accommodate events such as $D^{(*)}K^{(*)}\pi$,
and $D^{(*)}\rho$, where one or more pions or photons are missed in
the reconstruction and/or a pion is misidentified as a kaon.
The probability density function (PDF) is the sum of the signal and three
background components. A large $\Bz\ra\Dstarm\pip$ data control sample
is used to determine the signal shape in both \DE and \mES, and the
peaking background A in \DE, where we assign the kaon mass to 
the pion candidate. We use the same parameters for signal and peaking
backgrounds in \mES since they are consistent in simulation.
The \DE distributions and yields for the four components in the signal
region are shown in 
Fig.~\ref{fig:de} and Table~\ref{tab:par_data_full}, respectively.

\begin{table*}[htb]
\caption{The yields of signal, combinatorial (comb.) and
  peaking (peak A, peak B) background PDFs of the samples (a)--(d)
  described in the text;
  values and errors are 
  rescaled to represent the yields in the signal region
  ($\mES>5.27~\gevcc$, $|\DE|<40~\mev$). The bottom row shows the
  branching fractions with statistical errors. 
} 
\begin{center}
\begin{tabular*}{0.99\textwidth}{@{\extracolsep{\fill}}lr@{\,$\pm$}lr@{\,$\pm$}lr@{\,$\pm$}lr@{\,$\pm$}lr@{\,$\pm$}lr@{\,$\pm$}lr@{\,$\pm$}lr@{\,$\pm$}lr@{\,$\pm$}lr@{\,$\pm$}lr@{\,$\pm$}lr@{\,$\pm$}l}
\hline\hline
\bigstrut[t] & \multicolumn{6}{c}{ (a) $\Bz\ra\Dzb\Kp\pim$ }
 & \multicolumn{6}{c}{ (b) $\Bz\ra\Dz\Kp\pim$ }
 & \multicolumn{6}{c}{ (c) $\Bz\ra\Dzb\Kstz$ }
 & \multicolumn{6}{c}{ (d) $\Bz\ra\Dtwostm\Kp$ }\\
\Dz mode
 & \multicolumn{2}{c}{$K\pi$} & \multicolumn{2}{c}{$K\pi\piz$} & \multicolumn{2}{c}{$K\pi\pi\pi$}
 & \multicolumn{2}{c}{$K\pi$} & \multicolumn{2}{c}{$K\pi\piz$} & \multicolumn{2}{c}{$K\pi\pi\pi$}
 & \multicolumn{2}{c}{$K\pi$} & \multicolumn{2}{c}{$K\pi\piz$} & \multicolumn{2}{c}{$K\pi\pi\pi$}
 & \multicolumn{2}{c}{$K\pi$} & \multicolumn{2}{c}{$K\pi\piz$} & \multicolumn{2}{c}{$K\pi\pi\pi$} \\
\hline
Signal &  101 & 17  &  58 & 20 &  69 & 19 
       & $-17$ & 13 & 34 & 24 &    8 & 22 
       &  35 &  7   & 21 &  7 &  31  & 7 
       &  15  & 6 &  15  & 6 &   16  & 5\\
Comb.  &  229 &  4  & 500 &  5 & 528 &  5
       &  608 &  5 & 918 &  6 & 989 &  6 
       &  17 &  1 &  29 & 1  & 30 &  1 
       &  16 & 1  &  16  & 1 & 22 & 1 \\
Peak A &    5 &  6  &   0 &  1 &   0 &  2 
       &   0 & 0 & 0 & 0 & 0 & 0
       &   0 & 0 & 0 & 0 & 0 & 0
       &  2 & 2 & 5 & 2 & 2 & 1\\
Peak B &   45 &  9  & 76  & 12 &  42 & 10
       &  50 & 11  &  54  & 14 &  45 & 13
       &  6 & 3 & 10 & 3 & 3 & 3 
       &  2 &  3 &  7 & 3 & 0 & 1\\
\hline
\bigstrut[t] ${\cal B}$ $(10^{-6})$ & 
\multicolumn{6}{c}{$88 \pm 15 $}  &
\multicolumn{6}{c}{$-4\pm 12$ }  &
\multicolumn{6}{c}{$38\pm 6$}  &
\multicolumn{6}{c}{$18.3\pm 4.0$} \\
\hline\hline
\end{tabular*}
\end{center}
\label{tab:par_data_full}
\end{table*}

The signal yield for $\Bz\ra\Dzb\Kp\pim$ is
corrected for variations in signal efficiency across the
$DK\pi$ Dalitz plot.
Each event $k$ with variables $\vec{q}_k\equiv (m_{\mathrm{ES},k},\DE_k)$ is
assigned a signal weight~\cite{splot}
$$
w_\mathrm{sig}(\vec{q}_k)= \frac{\sum_{j=1}^{4}
  V_{\mathrm{sig},j}P_j(\vec{q}_k)} 
{\sum_{j=1}^{4} N_j P_j(\vec{q}_k)}\,,
\label{eq:sweight}
$$
calculated from the four PDF components $P_j$, their yields $N_j$
from the fit, and the covariance matrix elements $V_{\mathrm{sig},j}$
between $N_\mathrm{sig}$ and $N_j$.
The efficiency-corrected signal yield is then 
$\sum_k w_\mathrm{sig}(\vec{q}_k)/\varepsilon_k$, where the efficiency
$\varepsilon_k$ is estimated
from the simulated events in the vicinity of each data point in
the Dalitz plot.

\begin{figure}[tb]
\begin{center}
 \includegraphics[width=0.48\textwidth]{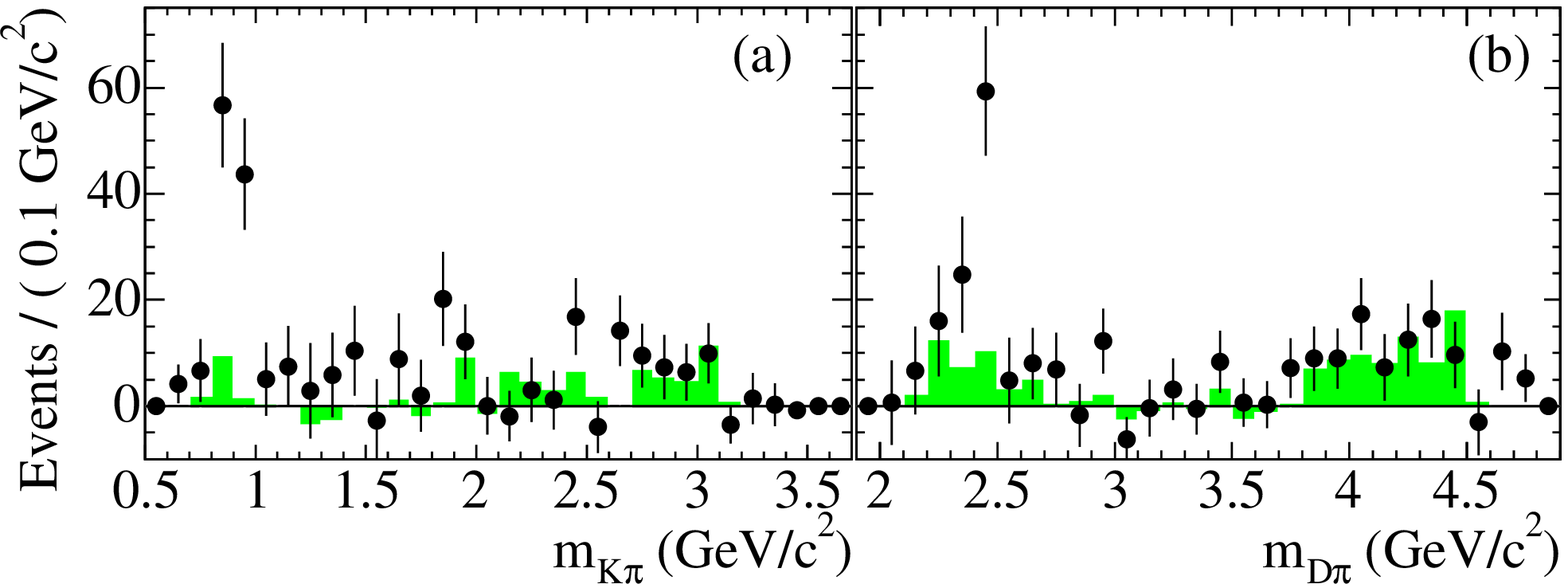}
\end{center}
\caption{The signal weight distribution as a function of $m_{\Kp\pim}$
  and $m_{\Dzb\pim}$. The shaded histograms include only events with (a)
  $|m_{\Dzb\pim}-2460~\mevcc|< 75~\mevcc$, and (b) $|m_{\Kp\pim}-896~\mevcc|<
  150~\mevcc$.}   
\label{fig:proj}
\end{figure}

Figure~\ref{fig:proj} shows the signal weight distribution as a function
of $m_{\Kp\pim}$ and $m_{\Dzb\pim}$. The peaks near $m_{\Kstz}$ and
$m_{\Dtwostm}$ are clearly visible. We use the $(\mES,\DE)$ fit results
and signal efficiencies estimated from simulated $\Bz\ra\Dzb\Kstz$ and
$\Bz\ra\Dtwostm\Kp$ samples
to compute corresponding branching fractions. 
For the $\Bz\ra\Dz\Kp\pim$ mode, we assume a flat distribution
on the Dalitz plot when determining the signal efficiency.

For modes in which we do not observe a significant signal, the 90\%
confidence level (C.L.)
branching fraction upper limit is determined by integrating the product
of the PDFs for the three \Dz modes as a function of
branching fraction from 
0 to ${\cal B}_\mathrm{UL}$ so that 
$\int_0^{{\cal B}_\mathrm{UL}} {\cal L} d{\cal B}= 0.9 \int_0^\infty
{\cal L} d{\cal B} $, where ${\cal L}$ is the likelihood function.

To measure $r$, we select events with $\mES>5.27~\gevcc$ from sample
(e).
A two-dimensional PDF of \DE and \thetaC (the reconstructed Cherenkov-light
angle of the prompt track) is used to 
separate $\Dstar K$ from $\Dstar \pi$ decays. 
Tracks with an estimated \thetaC uncertainty $\sigma_C > 4$~mrad or
$n_{\gamma,s}/\sqrt{n_{\gamma,s}+n_{\gamma,b}}<3$ are removed, 
where $n_{\gamma,s}$ and $n_{\gamma,b}$ 
are the numbers of signal and background photons determined from a
likelihood fit to the ring of Cherenkov photons associated with the
track~\cite{DIRC}.  
Finally events are rejected if \thetaC is smaller than the predicted
Cherenkov angle for kaons by more than $4\sigma_C$,
in order to remove particles heavier than kaon.

The \DE signal peak PDF is a Crystal Ball lineshape and the background
is a linear function plus a Gaussian peaked near $-150~\mev$ to
accommodate background events such as $\Dstar\rho$ and $D^{**}\pi$
where a soft $\pi$ is missed in the reconstruction.%
The distribution of $(\thetaC-\thetaC^\pi)/\sigma_C$ is modeled by
Gaussian functions. For the pion component, we use three Gaussian functions
centered near zero. For the kaon component, a single Gaussian
function centered near $(\thetaC^K-\thetaC^\pi)/\sigma_C$ is sufficient,
where $\thetaC^K$ and $\thetaC^\pi$ are the expected Cherenkov angle for
kaon and pion, respectively, based on the measured momentum.
Most of the parameters are obtained from a fit to the pion or kaon
tracks in a large $\ccbar\ra\Dstar X\ra\Dz\pi X,$ $\Dz\ra\Km\pip$ data control
sample, except the total width of the distribution, which is free in
the final fit to
accommodate a small difference in width due to differences in momentum
spectra between signal and control samples.

\begin{figure}[tb]
\begin{center}
 \includegraphics[width=0.48\textwidth]{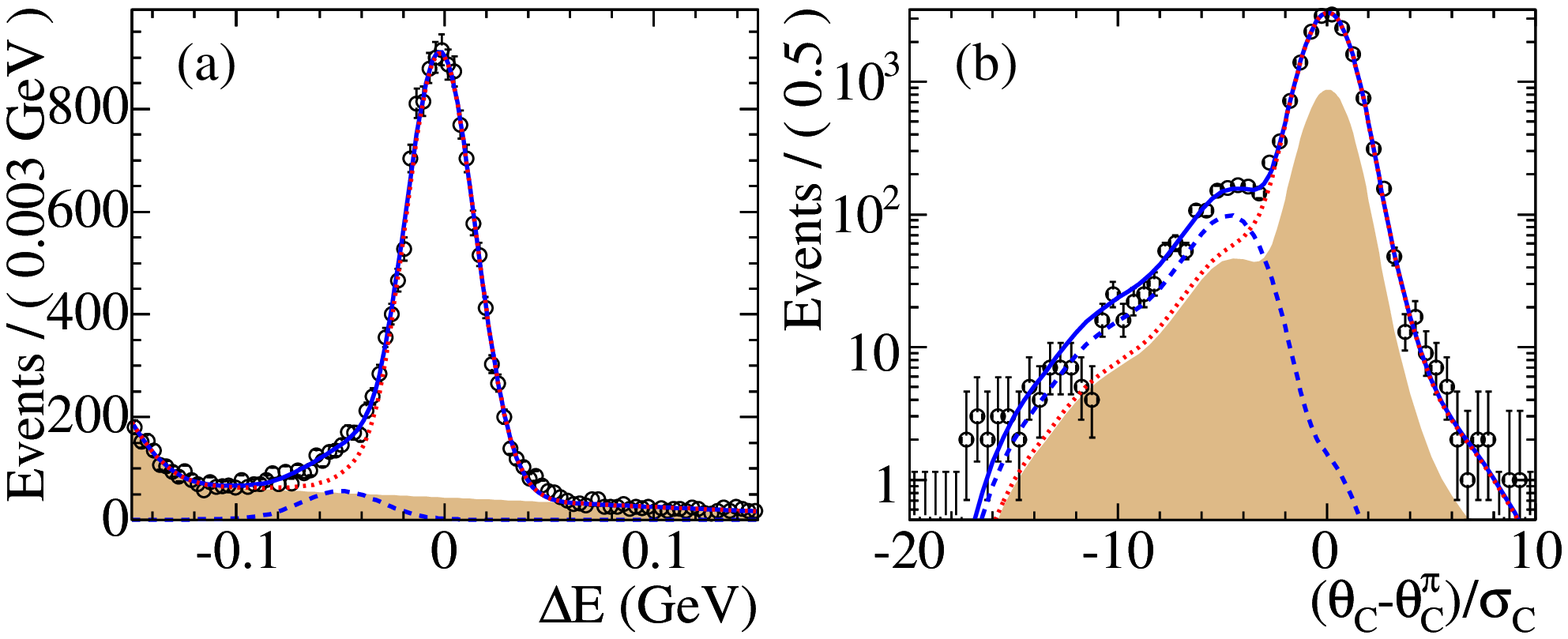}
\end{center}
\caption{(a) \DE and (b) Cherenkov angle $(\thetaC-\thetaC^\pi)/\sigma_C$
  distributions for $\Dstarm h^+$ candidates and PDF projections. 
  Circles with error bars are data points.
  Shaded distribution is combinatorial background, the dotted curve 
  adds the $\Dstar\pi$ contribution, and the solid curve is the full
  PDF. The dashed curve represents the $\Dstar K$ contribution only.
  \DE for $\Dstar\pi$ is centered near zero, while for
  $\Dstar K$ it is shifted to lower values because the prompt track is
  assumed to be a pion.} 
\label{fig:rdstk}
\end{figure}

Figure~\ref{fig:rdstk} shows the \DE and
$(\thetaC-\thetaC^\pi)/\sigma_C$ distributions and PDF projections for
$\Bz\ra\Dstarm h^+$ ($h = \pi$ or $K$) candidates.
We find 13400 signal events, of which $f= (6.80\pm 0.28)\%$  are
$\Dstar K$ events,
and 4850 background events in the sample.
The ratio $r= f/(1-f)$ is corrected by the signal efficiency ratio
$r_\varepsilon=\varepsilon_{\Dstar K}/\varepsilon_{\Dstar\pi}= (94.0\pm 2.3)\%$
obtained from simulation. 
This ratio is smaller than unity 
because \thetaC for kaons is smaller (resulting in fewer Cherenkov
photons) and more kaons than pions decay in flight within the tracking volume.
The uncertainty on $r_\varepsilon$ includes simulation statistics and
systematic uncertainties due to the two aforementioned effects.

For samples (a)--(d), the systematic uncertainties on the signal
efficiency are studied with 
large $\tau$ lepton decay samples (for track reconstruction efficiency) and
comparisons between signal simulation and 
$\Bz\ra\Dstarm \pip$ data control sample. 
The fractional uncertainty, common to all four samples, on signal
efficiency is 5\% including the uncertainties
on the number of \BB events and the \Dz branching fractions.
For the $\Bz\ra\Dzb\Kp\pim$ mode, the uncertainty of efficiency
variation on the Dalitz plot
contributes an additional systematic error of 8\%.
In addition, we vary the control sample shapes in each fit by one standard
error and sum the changes in signal yield in quadrature. The total signal yield
variations are 8, 2.0, 3.4, and 2.6 events
for $\Dzb\Kp\pim$, $\Dz\Kp\pim$, $\Dzb\Kstz$, and $\Dtwostm\Kp$,
respectively. 
For the $\Bz\ra\Dzb\Kstz$ and $\Dtwostm\Kp$ measurements, we consider
possible contamination from each other and from the non-resonance
contribution. 
Using the signal yields for $\Bz\ra\Dzb\Kstz$ and $\Dtwostm\Kp$, and
the cross-feed efficiencies
determined from simulation, we find that six events in each of these
two \Bz modes could be attributed to the other mode and to non-resonance
contributions. This contributes a 6\% uncertainty for
$\Bz\ra\Dzb\Kstz$ and 11\% for $\Bz\ra\Dtwostm\Kp$.
The uncertainty due to the full width of the \Dtwostm and \Kstz resonances is 
8\% for $\Bz\ra\Dtwostm\Kp$ and less than 1\% for $\Bz\ra\Dzb\Kstz$.

The largest systematic uncertainties cancel in the branching ratio measurement
(sample (e)).
The remaining systematic errors are from PDF shapes, control sample
distributions and contaminations ($1.9\%$), 
residual uncertainties in the signal efficiency ratio ($2.4\%$), and
potential fit bias ($2.1\%$). The last item has been evaluated
with simulation samples including background.

In conclusion, we have measured the branching fraction for the
$\Bz\ra\Dzb\Kp\pim$ decay excluding $\Dstarm\Kp$,
$$
{\cal B}(\Bz\ra\Dzb\Kp\pim) = (88 \pm 15 \pm 9 )\times 10^{-6},\\
$$
as well as its two significant resonances,
\begin{gather*}
{\cal B}(\Bz\ra\Dzb\Kstz)\cdot {\cal B}(\Kstz\ra\Kp\pim) \\
=  (38\pm 6 \pm 4) \times 10^{-6}\,,\;\;\;\;\;\text{and} \\
{\cal B}(\Bz\ra\Dtwostm\Kp)\cdot {\cal B}(\Dtwostm\ra\Dzb\pim)\\
= (18.3\pm 4.0 \pm 3.1) \times 10^{-6}\;.
\end{gather*}
The signal significances are 8.7, 8.3 and 5.0
standard deviations, respectively, determined from the change in the
likelihood between the best fit 
and a fit with the signal yield fixed to zero (the first case) or the
possible cross feed from other sources (six events for the latter
two cases).
From a fit excluding the observed resonances, assuming flat
distriubtion on the Dalitz plot, we find 
${\cal B}(\Bz\ra\Dzb\Kp\pim)= (26\pm 8\pm 4)\times 10^{-6}$,
whose signal significance is $3.1\sigma$ and 
90\% confidence level upper limit is $37\times 10^{-6}$. 
We do not observe a significant signal for 
the CKM-suppressed $\Bz\ra\Dz\Kp\pim$ mode.
The 90\% confidence level upper limit is 
${\cal B}(\Bz\ra\Dz\Kp\pim) < 19\times 10^{-6}$.
The event yields in this channel are lower than
anticipated~\cite{APS} indicating that a significantly larger data
sample is required 
to constrain $\gamma$ through this method.

The ratio of branching fractions for $\Bz\ra\Dstarm\Kp$ to
$\Bz\ra\Dstarm\pip$ is measured to be
$$
r= (7.76\pm 0.34 \pm 0.29)\%\;,
$$
a nearly four-fold improvement compared to the
previous result~\cite{Abe:2001wa}.
This ratio is consistent with $(f_K/f_\pi)^2\tan^2\theta_\mathrm{Cab} \simeq
0.072$~\cite{decayconstant}, expected at tree level if factorization
and flavor-SU(3) 
symmetry hold, where $\theta_\mathrm{Cab}$ is the Cabibbo angle and $f_K$ and
$f_\pi$ are the decay constants of the kaon and pion, 
respectively.

We are grateful for the excellent luminosity and machine conditions
provided by our \pep2\ colleagues, 
and for the substantial dedicated effort from
the computing organizations that support \babar.
The collaborating institutions wish to thank 
SLAC for its support and kind hospitality. 
This work is supported by
DOE
and NSF (USA),
NSERC (Canada),
IHEP (China),
CEA and
CNRS-IN2P3
(France),
BMBF and DFG
(Germany),
INFN (Italy),
FOM (The Netherlands),
NFR (Norway),
MIST (Russia), and
PPARC (United Kingdom). 
Individuals have received support from CONACyT (Mexico), A.~P.~Sloan Foundation, 
Research Corporation,
and Alexander von Humboldt Foundation.



\begin{thebibliography}{99}
\bibitem{footnote1}
The charge conjugate state is implied throughout this Letter.
\bibitem{crball}
The Crystall Ball lineshape is a modified Gaussian distribution with a
transition to a tail function on one side
$[(n/\alpha)^n\exp(-\alpha^2/2)]/((\overline{x}-x)/\sigma+n/\alpha-\alpha)^n$
when $x\le \overline{x}-\alpha\sigma$, where $\overline{x}$ and
$\sigma$ are the mean and width of the Gaussian for $x >
\overline{x}-\alpha\sigma$.
\bibitem{decayconstant}
The value is calculated from
$\Gamma(\taum\ra\Km\nu_\tau)/\Gamma(\taum\ra\pim\nu_\tau)$, corrected
for phase space factors.
\end{thebibliography}
\end{document}